# Surface phase transitions in foams and emulsions

Nikolai Denkov*, Slavka Tcholakova, Diana Cholakova

*Department of Chemical and Pharmaceutical Engineering*
*Faculty of Chemistry and Pharmacy, Sofia University,*
*1 James Bourchier Avenue, 1164 Sofia, Bulgaria*

*Corresponding author:
Prof. Nikolai Denkov
Corresponding member of the Bulgarian Academy of Sciences (BAS)
Department of Chemical and Pharmaceutical Engineering, Sofia University
1 James Bourchier Ave., Sofia 1164
Bulgaria
E-mail: nd@lcpe.uni-sofia.bg
Tel: +359 2 8161639
Fax: +359 2 9625643

# Abstract

Surface phase transitions in surfactant adsorption layers are known to affect the dynamic properties of foams and to induce surface nucleation in freezing emulsion drops. Recently, these transitions were found to play a role in several other phenomena, opening new opportunities for controlling foam and emulsion properties. This review presents a brief outlook of the emerging opportunities in this area. Three topics are emphasized: (1) The use of surfactant mixtures for inducing phase transitions on bubble surfaces in foams; (2) The peculiar properties of natural surfactants saponins which form extremely viscoelastic surface layers; and (3) The main phenomena in emulsions, for which the surface phase transitions are important. The overall conclusion from the reviewed literature is that surface phase transitions could be used as a powerful tool to control many foam and emulsion properties, but we need deeper understanding of the underlying phenomena to explore fully these opportunities.

**Key words:** surface phase transition, liquid condensed state, liquid expanded state, adsorption layer, surfactant mixture, foam, emulsion, partial coalescence, surface nucleation, drop self-shaping, saponin.

## 1. Introduction – role of surface "rigidity" in foam studies.

The role of surface "rigidity" on the mode of foam film thinning was first described by Miles et al. [1] and then elaborated in great detail in the classical book "Soap Films" by Mysels et al. [2]. In a series of experimental observations with solutions of different surfactants, these authors revealed several modes of foam film thinning and the terms "rigid" and "mobile" films were introduced for the two extreme types of behaviour. Epstein et al. [3] pointed out that the surface rigidity is probably due to formation of "solid mixed surface layers" and emphasized the role of the dodecyl alcohol ($C_{12}OH$) as an essential admixture which rigidifies the adsorption layers of sodium dodecyl sulfate (SDS) which was used as main surfactant in most experiments. Thus, the intricate link between the phase state of surfactant adsorption layers and the film thinning behaviour was recognized for the first time.

In the following years, several groups developed theoretical approaches to quantify the relations between the viscoelastic properties of the adsorption layers and the foam film thinning behaviour. Ivanov and collaborators [4] analysed in detail the role of surfactant mass transfer and the related Marangoni effect which creates apparent surface elasticity. Langevin [5] analysed the role of surface viscosity. Joye et al. [6] explained the role of surface viscoelasticity in the phenomenon of marginal regeneration observed in thinning foam films. Somewhat surprisingly, neither of these studies tried to establish a deeper connection between the film thinning processes and the wide research area of surface phase transitions in spread surfactant layers, which was well established in the literature by that time, see Section 2 below.

It took almost 40 years after the book "Soap Films" [2] was published, when it became clear that the surface phase transitions have a profound impact on many other dynamic properties of foams, see **Figure 1**. First, in his studies of the coalescence stability of foam films, Bergeron [7] showed that the surface dilatational elasticity might be an equally important factor for the film stabilization, along with the well-known surface forces (electrostatic, van der Waals, etc.).

Second, an inspiring debate about the rate of liquid drainage from foams [8,9] ended with the conclusion that two different regimes are possible, depending on the surface mobility of the foam bubbles. By adding lauryl alcohol to SDS solutions, Stoyanov et al. [10] changed the regime of liquid drainage, thus demonstrating the direct link to the surface phase transitions. Theoretical models were developed to describe the regimes of "rigid" and "fully mobile" interfaces [11,12].

Third, in a series of papers Denkov et al. [13,14] showed that the foam rheological response also exhibits two qualitatively different regimes, controlled by the surface mobility of the bubbles. Notably, it was shown for the first time in these studies [13-15] that the oscillating drop method is a very convenient tool to characterize the surface mobility. High surface moduli $E > 100$ mN/m, as measured at low oscillation frequency (0.1 to 1 Hz), always corresponded to the regime of "rigid" interfaces, whereas $E < 50$ mN/m always corresponded to mobile interfaces [15]. The long-chain fatty acids (C14 to C18) were introduced in these studies as convenient cosurfactants to induce surface phase-transition (condensation).

In parallel series of experiments, Langevin et al. [16,17] and Tcholakova et al. [18] studied the role of surface elasticity on the bubble Ostwald ripening in foams. Both series of experiments concluded that the bubble ripening is much slower at high surface elasticity. Two complementary explanations were given to this effect [16-18]. Both the surface elasticity and the reduced gas permeability of the condensed adsorption layers could play a role for the strong reduction of the Ostwald ripening rate, and the relative importance of these two effects is system-specific.

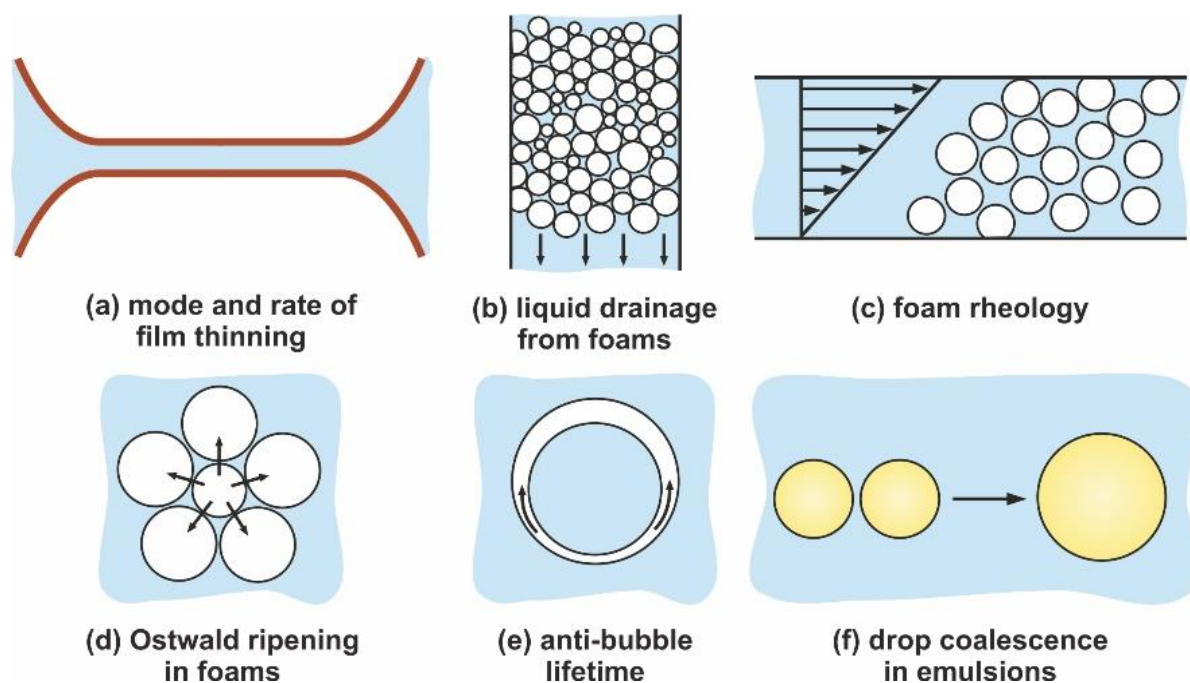

**Figure 1.** Schematic presentation of several phenomena for which the surface phase transitions in the surfactant adsorption layers change very significantly the foam or emulsion behaviour.

The various effects mentioned above have been studied multiple times in the following years by various research groups which confirmed the important role of surface condensation in all dynamic properties of the foams. Thus, the role of surface rigidity/mobility in foams has been firmly established. Nevertheless, a systematic study about the relation between the foam properties and the various possible surface phases in the adsorption layers (Section 2a) has not been made so far.

The role of surface phase transitions in the adsorption layers covering the oily drops was also studied by several groups in the context of drop crystallization and drop-drop partial coalescence in cooled oil-in-water emulsions [19-21].

The major aim of this review is to summarize the recent progress in the area of surface phase transitions in foams and emulsions. The important open questions and possible directions for future research are outlined.

Due to the limited space, the focus of this review is exclusively on the adsorption layers of low-molecular-mass surfactants (incl. saponins). The adsorption layers of solid particles and polymer molecules (including the biopolymers, such as proteins and polysaccharides) exhibit a large variety of additional, strongly system-specific effects which are not reviewed here.

## 2. Surface phase transitions at air-water interface in relation to foam properties.

### *(a) Main surface transitions and experimental methods for their investigation.*

It is accepted to call "Langmuir monolayers" the spread surface monolayers of insoluble molecules on a liquid sub-phase. In contrast, the monolayers formed via spontaneous adsorption of dissolved amphiphilic molecules from adjacent bulk phases are called "Gibbs adsorption layers" [22]. For brevity, we also use this notation below.

Historically, the surface phase transitions were first observed in Langmuir monolayers. Using Langmuir trough to vary precisely the area-per-molecule in the layer and measuring the surface tension by Wilhelmy plate method allowed the researchers to detect the main surface phase transitions and to clarify the role of the main factors involved. Typically, the surface phase transitions follow (part of) the sequence gas → liquid expanded → liquid condensed → solid (G → LE → LC → S), as the temperature and/or the area-per-molecule is reduced, see **Figure 2** [23-25]. The specific temperatures and adsorptions, at which these surface phase transitions are observed, depend strongly on the head-group, chain length, branching and presence of double bonds in the surfactant tail. Summary of these classical studies can be found, e.g., in the books by Gaines [23] and Adamson & Gast [24].

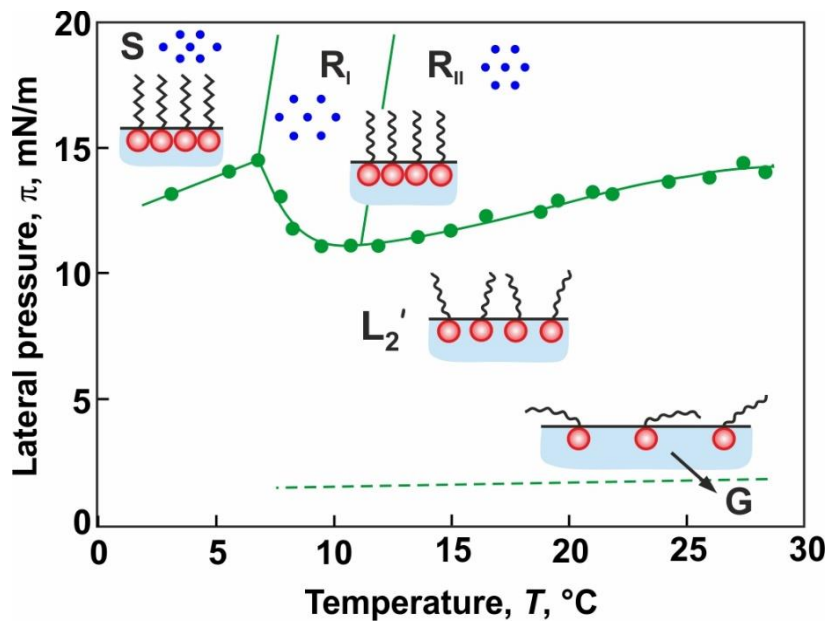

**Figure 2.** Schematic presentation of the main surface phases, observed in octadecanol monolayers spread on water surface. G denotes gas phase, $L_2'$ - liquid expanded (disordered) phase, $R_I$ and $R_{II}$ – two liquid condensed (rotator) phases, S – crystalline solid phase. The various phases differ in the area-per-molecule, tilt of the alkyl chain, and in-plane structure, as shown in the figure (adapted from [25]).

More recently, many powerful structural methods were developed or adapted for surface studies and they revealed a rich variety of polymorphic structures formed at the air-water interface. The invention of monolayer-sensitive microscopy techniques, such as fluorescence microscopy, Brewster-angle microscopy [26] and polarized fluorescence microscopy [27] revealed mesoscopic structures of peculiar shapes in Langmuir monolayers. The fine molecular structure of the surface layers was characterized by synchrotron techniques, including X-ray scattering, X-ray reflectivity, Grazing incidence diffraction (GIXD), neutron scattering and reflectivity, as well as by various spectroscopy techniques, such as surface-specific sum-frequency generation spectroscopy and infrared reflection-absorption spectroscopy [28]. Atomic force microscopy of Langmuir-Blodgett layers transferred onto solid substrates was also used to analyze the structure of Langmuir monolayers and of more complex surface aggregates and multilayers [29].

During the last years, surface rheological measurements by the oscillating drop method [30], capillary pressure tensiometer [31], magnetic needle [32,33] and double wall-ring method [34] provided valuable complementary information about the phase transitions in Langmuir and Gibbs layers. For example, a significant increase in the surface viscosity was observed upon phase transition LE → LC in tri-component DPPC:PA:Chol monolayers, studied at different surface pressures and cholesterol fractions [35**]. Recent reviews on the methods for characterization of the surface rheological properties are presented in [36,37].

It is worth mentioning that several recent studies [38-40] warned about possible problems with some of these experimental methods. Thus, it was shown that Langmuir trough-based techniques may give unreliable results for the reversibility of the surface phase transitions, when reaching low surface tensions, due to the leakage of material [38]. An approach to overcome this problem was proposed in [39] where reversible phase transition in phospholipid layers was studied by a new method, called "constrained drop surfactometry", which is claimed to provide a leakage-proof environment.

In another study, Cuenca et al. [40] resolved the previous puzzling reports about the measurements of apparently negative surface viscosity in some surfactant systems. The authors found that the surfactant adsorption-desorption processes might be highly asymmetric, with different characteristic times and mechanisms. This asymmetry invalidates the accepted conventional procedure for data analysis that leads to physically unrealistic negative values of surface viscosity.

Because the Langmuir layers of insoluble surfactants are inefficient for stabilizing foams and emulsions, below we focus our review on the Gibbs layers, formed via spontaneous adsorption of surfactant molecules from the adjacent surfactant solution. Several approaches have been applied to induce phase transitions in Gibbs monolayers, as illustrated in **Figure 3** and described below.

### (b) *Mixtures of soluble and insoluble surfactants*

The solubilization of water-insoluble surfactants in the micelles of water-soluble surfactants has proven to be a powerful approach for forming mixed adsorption layer which can undergo surface phase transitions with the formation of highly viscoelastic surfaces, see **Figures 3a** and **4**. Usually fatty alcohols or fatty acids are used as cosurfactants to various anionic or cationic main surfactants [15-17,41-43*]. Several recent studies are reviewed below to illustrate the universal character, the versatile use, and the importance of this approach.

In [42] it was shown that the addition of $C_{12}OH$ to SDS-containing foaming solutions increases significantly the foam stability, due to the increased viscoelasticity of the formed mixed adsorption layers. In contrast, the foam stability is governed mostly by electrostatic interactions (and is lower) at the same concentration of SDS, in the absence of $C_{12}OH$. These results are another convincing support to the concept [7] that the surface viscoelasticity plays a very important role in foam stabilization.

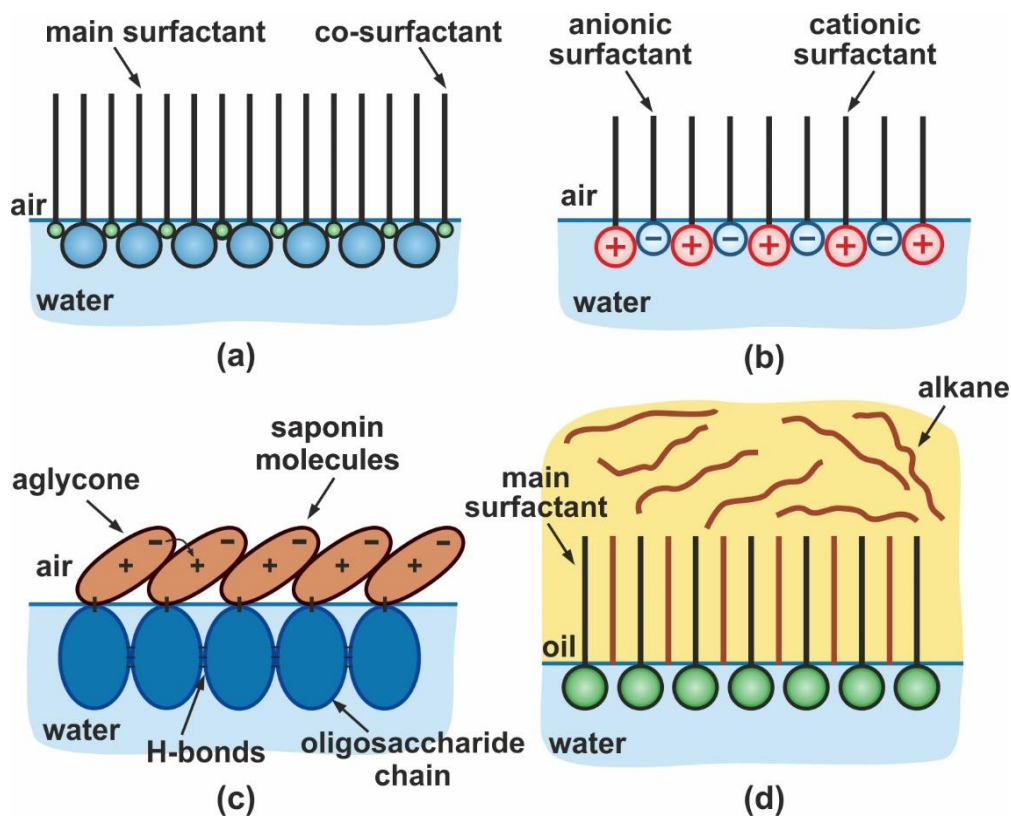

**Figure 3.** Schematic illustration of the main approaches used to induce surface phase transition in adsorption layers of low-molecular-mass surfactants at air-water and oil-water interfaces. (a) Addition of poorly water-soluble cosurfactants with small head-groups. The cosurfactants are solubilized in the surfactant micelles and form mixed adsorption layers with the main surfactants at the air-water interface. (b) Mixing anionic and cationic surfactants in appropriate ratio so that surface complex of the two surfactants is formed. Alternatively, multivalent counterions could be used to induce condensation of the adsorption layer of charged surfactant molecules and even to form surface multilayers of these molecules. (c) Saponin molecules pack well on air-water and oil-water interfaces, due to strong attraction between their large hydrophobic aglycons and to multiple hydrogen bonds between their hydrophilic sugar groups. (d) Interdigitating long-chain alkane molecules into adsorption layers of long-chain surfactants leads to formation of mixed alkane-surfactant monolayers which may freeze at temperatures above the freezing temperature of the adjacent bulk phases.

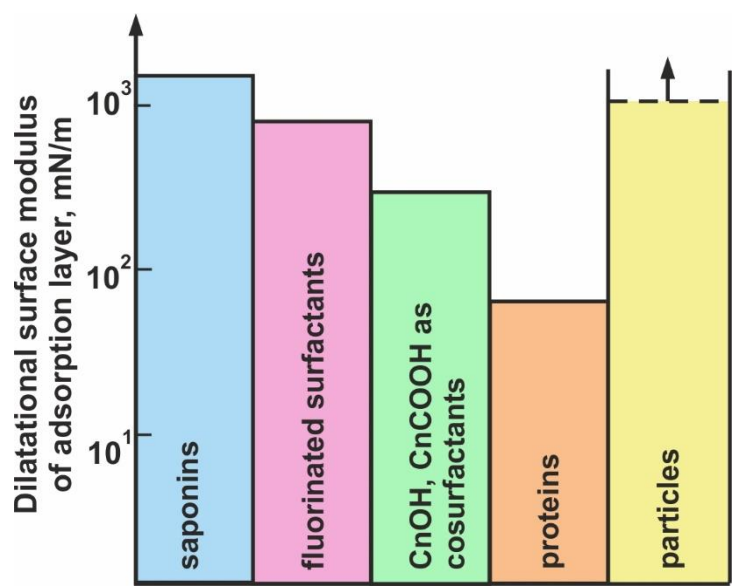

**Figure 4.** Typical range of the dilatational surface moduli of adsorption layers of water-soluble low-molecular-mass surfactants, as measured at low oscillation frequencies (ca. 0.1 to 1 Hz) in comparison with adsorption layers of common proteins and with particle interfacial layers.

In a recent study with mixed solutions of myristic acid (MAc) and choline hydroxide [44*] it was shown that the pH and the ratio of the two components play a key role for foam stabilization. The choline is used as a bulky counter-ion which increases the solubility of MAc at low pH. The authors found that the foam stability has a maximum at pH values around the pK of MAc. Under these conditions, dense mixed adsorption layers of ionized and nonionized species of MAc are formed at the bubble surface and stabilize the foam very efficiently. In other words, the adsorption layer behaves as composed of two different surfactant components, one of them being a fatty acid with small head-group, while the other is the ionic form of this acid. Thus, the layer structure closely resembles the one shown in **Figure 3a**, despite the fact that only one surfactant substance is dissolved initially.

In a systematic series of experiments, the effects of chain-length and concentration of fatty acids (used as cosurfactants), pH and temperature on the onset of surface phase transition and on the related foam behavior were studied by Mitrinova et al. [15,43*]. Several important conclusions were drawn: (1) Most viscoelastic layers at 21 °C were obtained with myristic and palmitic acids having chain-lengths of C14 and C16, respectively. The longer stearic acid with 18 carbon atoms was not solubilized well in the micelles of the main surfactants and, hence, the viscoelasticity of the respective mixed adsorption layers was lower as compared to the myristic and palmitic acids; (2) Increasing temperature and/or pH can be used to melt the surface layers which leads to jump-wise decrease of the surface viscoelasticity by ca. 2 orders of magnitude; (3) neutron reflectivity measurements showed that (a) the cosurfactant molar fraction in the adsorption layer is comparable to that of the main surfactant, despite the much lower cosurfactant concentration in the bulk solution, and (b) the viscoelastic layers correspond to an average area per molecule of ≈ 0.20 nm$^2$ (unpublished results). This latter value evidences that the molecules in the viscoelastic layers have area-per-molecule resembling that in the intermediate rotator phases, observed with bulk alkanes, rather than the area corresponding to fully crystalline structure of alkyl chains which is < 0.19 nm$^2$ [45]. (4) Well-defined boundary exists in the surface viscoelastic modulus, between 50 and 100 mN/m (depending on the specific foam phenomenon studied), which leads to qualitative shift in the foam behavior and thus separates the regimes of rigid and fluid interfaces.

Marinova et al. [46] applied this approach to even more complex mixtures of surfactants, anionic SLES + nonionic APG + zwitterionic CAPB. The addition of $C_{12}OH$ was found to decelerate the rate of water drainage from the respective foams at high salt concentration and low temperature. These results were explained by the higher viscoelasticity of the $C_{12}OH$-containing adsorption layers.

Several papers reported that the surface viscoelasticity plays a role in foam systems relevant to enhanced oil recovery. Thus, Wang et al. showed [47] that the addition of long chain fatty acids which increase the surface viscoelasticity of adsorption layers could strongly enhance the foam tolerance to the antifoam action of oils. In an independent study, Mensire et al. [48] showed that the rheological properties of the interfaces and of the respective foams affect strongly the oil extraction parameter from porous rocks. Along the same line of reasoning, Wang et al. [49,50] showed that higher surface dilatational modulus decreases the surface deformability and, thus, impacts the foam flow in porous medium.

Several studies showed that the surface viscoelasticity is very important for other foam-related phenomena as well. Dollet and Bocher [51] showed that the surface viscoelasticity changes qualitatively the dynamic behavior of a bubble monolayer (2D foam) flowing between two solid plates. These results were explained by considering the role of surface viscoelasticity on the bubble-wall friction, bubble elastic deformation, and the rate of bubble plastic rearrangement. Vitry et al. [52] showed that the surface elasticity is a key factor to increase the lifetime of the so-called "antibubbles" - spherical gas films, freely floating in surfactant solutions, **Figure 1e**.

### *(c) Mixtures of oppositely charged surfactants (catanionic surfactants) and of ionic surfactants with multivalent counterions.*

In a series of papers, Langevin and co-workers studied the foaming properties of mixtures of CTAB and myristic acid, considered as composed of oppositely charged surfactants. In [16,17] they showed that mixed adsorption layers with low surface tension and high surface elasticity can be formed. The respective foams were very stable to bubble Ostwald ripening and bubble-bubble coalescence. The surface properties of these solutions were characterized in detail in [53*]. The existence of a melting temperature was reported above which the surface layer was fluidized, similarly to the observations in [15,43*].

As summarized in the very informative review by the same research group [54*], the formation of such highly viscoelastic surface layers, using mixtures of oppositely charged surfactants, is often combined with other mechanisms leading to ultra-stable foams, such as gelation of the foam films and/or of the bulk foaming solution. Recent example of such approach was presented in [55] where precipitation of the surfactant on the bubble surface was used to form particle-stabilized (Pickering) ultra-stable foams, responsive to temperature variations.

Lyadinskaya et al [56**] showed that the surface tension of CTAB, DTAB and DNA solutions does not change, if these solutes are dissolved separately at very low concentrations. However, if DNA and cationic surfactants are dissolved together at the same concentrations, strong synergistic effect is observed and the surface tension decreases due to the formation of DNA+surfactant complexes. The surface elasticity of the mixed DNA+CTAB solution increases up to 85 mN/m, while the maximal value for CTAB alone is ≈ 20 mN/m. The authors explained their results with combined strong electrostatic and hydrophobic attraction between the DNA and surfactant molecules which leads to surface complexation. The hydrophobic attraction is between the hydrophobic surfactant tail and the DNA nucleobases exposed in the major groove of the double helix. Several other interesting phenomena were described in these systems, such as transition from loose network to dense and brittle adsorption layer, formation of adsorption multilayers, and growth of fibrillar aggregates which can merge and form continuous surface phase [56**].

Another approach to induce surface phase transitions using appropriate control of the electrostatic interactions is to add multivalent counterions to ionic surfactants, which can lead to surface ordering and even to formation of surfactant multilayers at the interface. Such multilayers were detected recently by neutron reflectivity upon addition of multivalent cations to solutions of the anionic surfactants α-methyl ester sulfonate (MES) [57-59] and SLES [60].

*(d) Saponin adsorption layers and saponin-stabilized foams.*

Saponin is a generic term for a large class of natural amphiphilic molecules which are widely spread in the plant kingdom and can be extracted at industrial scale from many plant species. Some of the saponin extracts form highly viscoelastic adsorption layers at air-water interface and could be very efficient foam stabilizers. Therefore, studies on the saponin surface and foaming properties appear regularly during the last years.

The high viscoelasticity of the saponin adsorption layers is caused by strong attraction between the adsorbed saponin molecules which leads to surface phase transition, as evidenced by analysis of surface tension isotherms of saponin solutions [61*]. Molecular dynamic simulations [62*] showed that the intermolecular attraction includes strong dipole-dipole interactions and hydrogen bonding. Some saponins form elastic "skin" on the air-water and oil-water interfaces with strongly non-linear response upon surface deformation [63,64].

It is worth noting that the commercial saponin extracts are usually multicomponent mixtures, containing molecules with different solubility in water [61*,65,66]. Therefore, one expects that the formation of viscoelastic adsorption layers for many of these extracts could be affected by the presence of both highly water-soluble components (which form micelles) and poorly water-soluble components (which could be solubilized inside the micelles). These different components could form mixed adsorption layers with high viscoelasticity, just as the more conventional surfactant mixtures discussed in Section 2b.

Wojciechowski [65] and Golemanov et al. [66] clarified that the surface properties of the saponin solutions may differ significantly, depending on their source and producer. Kezwon and Wojciechowski [67*] studied also mixed adsorption layers of Quillaja saponin (QS) with proteins and lipids and strong surface complexation was reported for some of these systems. In a separate study, Botcher et al [68] showed that QS and the globular milk protein beta-lactoglobulin also form complexes on the air-water interface with strong synergistic effect on foam stability.

Ulaganathan et al. [69] studied the surface and foam film properties of solutions of QS. The effect of pH was found to be relatively small for the surface properties and for the rate of film thinning. The final film thickness, however, was affected by pH – an effect attributed to changed electrostatic repulsion between the foam film surfaces.

Pagureva et al [61*] found a clear relation between the surface viscoelasticity of the saponin adsorption layers and the rate of bubble Ostwald ripening in the respective foams. Two properties of the adsorption layers were found to be important for the decelerated bubble ripening in saponin-stabilized foams. First, the condensed saponin adsorption layers have very low permeability for the gas molecules which should cross the foam films in the process of gas exchange between the bubbles, **Figure 1d**. Second, the surface tension of the shrinking bubbles could be significantly lower than the equilibrium one, thus decreasing the bubble capillary pressure which is the main driving force for the gas transfer. These two combined effects reduce the rate of bubble Ostwald ripening by up to 2 orders of magnitude, as compared to the foams stabilized by conventional surfactants.

The foamability and foam density for solutions of different saponins were compared in [70] and [71]. However, these authors could not find clear correlations between the foam properties and the measured surface characteristics.

Enhanced foam stability after addition of chitosan to saponin solutions was reported in [72], especially with respect to the suppressed bubble coalescence during foaming. This effect was partially explained by the increased viscosity of the mixed saponin-chitosan solutions which leads to the formation of smaller and more stable bubbles.

Two new extracts of saponins from the plant species *Quillaja brasiliensis* and *Agave angustifolia* were studied in [73]. The authors found that the surface and foaming properties of the studied saponin solutions depended strongly on the method of extract preparation, in agreement with previous studies.

Stable foams were formed by solutions of glycyrrhizic acid (GA) in which saponin nanofibrils were formed and stabilized the bubbles [74].

One sees from the above brief review of the recent publications that the saponins are of high research interest. However, the complexity of their surface properties and the variability of the studied saponin extracts make very difficult the physicochemical interpretation of the obtained results. One could expect a slow but steady progress in this area.

*(e) Fluorinated surfactants*

Several recent studies showed that the fluorinated surfactants could be very efficient in forming highly viscoelastic adsorption layers and in enhancing foam stability. Thus, Kovalenko et al. [75**] showed that fluorinated surfactants, with phosphonic head-groups and 8 to 10 carbon atoms in their chains, form adsorption layers with very high elasticity, up to 950 mN/m. High elasticities, up to 250 mN/m, were reported also in [76*] for adsorption layers of heptadecafluoro-1-nonanol ($C_9H_2F_{17}OH$). Phase transition in the adsorption layer of the type LE → LC was reported and the high surface viscoelasticity was attributed to the LC phase. An independent study of the short-chain zwitterionic fluorocarbon cosurfactant FS50, added to SDS solutions, showed very strong synergistic effects and enhanced foam stability for the mixed solutions [77].

Obviously, the foaming properties of the fluorinated surfactants and their mixtures with the common alkyl-based surfactants are poorly understood, as we do not know the detailed mechanisms governing these surface phase transitions and leading to the reported very high surface viscoelasticities. New interesting results could be expected in this emerging research area.

### *(f) Stimuli responsive surfactants and smart foams*

There are rapidly growing interest and research activity in stimuli-responsive foams and emulsions. Temperature and pH were successfully used to modify in a desired way the surface and foaming properties of various surfactant solutions in a number of studies [78-82].

A new approach for foam control was developed recently which requires specially synthesized photo-sensitive surfactants with cis-trans isomerization of the surfactant molecules and relies on the surface phase transitions triggered by light illumination. Chevallier et al. [79-81**] characterized the surface and foam film properties for solutions of such light-sensitive surfactants. Several unexpected phenomena were observed due to non-trivial Marangoni effects and to changes in the surfactant aggregates when the surfactant conformations were shifted by the light illumination [79-81**].

Independently, Jiang et al. [82] showed that the foaming ability of the solutions of AZO-B4 light-sensitive surfactant is comparable to that of $C_{16}TAB$ and that the solution properties could be controlled using UV-illumination and/or pH variation of the solution, e.g. via $CO_2$ purging through the solution.

Certainly, this area will be expanded in the following years, because the stimuli responsive systems will remain of high interest from both academic and applied viewpoints.

## 3. Phase transitions at oil-water interface and their effect on emulsion properties.

One could expect that, similarly to foams, the phase transitions in the surfactant adsorption layers, formed at oil-water interfaces, should affect significantly the emulsion properties. Until recently, however, these effects have not been systematically studied, with one exception – in relation to the partial coalescence of the oily drops in cooled oil-in-water emulsions, when the drops undergo liquid-to-solid phase transition.

Arguably, the main reason for this limited interest in the surface phase transitions in emulsions is that the central approach to induce surface phase transitions in foams (namely via addition of water-insoluble cosurfactants with small head-groups, **Figure 2a**) is not applicable to emulsions. Indeed, the long-chain fatty alcohols and acids are highly soluble in the oily phase. Therefore, no condensed (frozen) adsorption layers are formed in emulsions under conditions which would produce such layers in foams.

Two other main approaches were found useful to induce surface phase transition at the oil-water interfaces with possible application to emulsions, as briefly explained below

### (a) Long-chain $C_n$TAB and $C_n$TAC surfactants.

The first approach relies on using long-chain trimetylammonim bromides or chlorides ($C_n$TAB or $C_n$TAC with $n$ = 14 to 18) [83-86*]. The relatively large charged head-group of these surfactants suppresses their solubility in the oily phase. Thus, $C_n$TAB or $C_n$TAC have very high affinity to the oil-water interface and readily form relatively dense adsorption layers. Interfacial freezing of these layers was observed only when alkanes with chain-length comparable to the surfactant tail were present in the hydrocarbon phase (as single constituents or in alkane mixtures). The experiments evidenced that the frozen adsorption layers contain surfactant and interdigitated alkane molecules which nicely pack together upon layer freezing [86*], see **Figure 3d**.

The series of papers by Bain, Aratono, Deutch, Ocko and their collaborators [84-88] revealed the role of various factors in the interfacial phase transitions at oil-water interface and opened the door for studying their effect on the emulsion properties. Very comprehensive review on the progress in this area was published recently by H. Matsubara and M. Aratono [89**].

Following this line of research, Tokiwa et al. [90*] demonstrated that alkane-in-water emulsions, stabilized by CTAC, remained stable for a period of 24 h, when stored at temperatures corresponding to frozen adsorption layer on the drop surfaces. In contrast, above the layer melting temperature, the emulsion became unstable - bulk oil phase separated immediately after emulsification and the emulsion volume decreased continuously with time due to drop-drop and drop-homophase coalescence. The authors explained this trend with the higher mechanical elasticity of the frozen adsorption layers and the related slower drainage of the emulsion films, formed between the neighboring droplets [90*].

### *(b) Long-chain monoglycerides and polysorbates in emulsions undergoing partial coalescence.*

The second approach was proposed in the studies of the partial coalescence phenomenon in cooled emulsions, **Figure 5**. It includes the use of long-chain monoglycerides or polysorbates (Tweens or Spans) which may induce interfacial freezing in such emulsions.

The partial coalescence occurs in the temperature range, in which the dispersed oily phase is partially crystallized [91]. The process is triggered by the solid crystals which appear in the droplets upon emulsion cooling [92,93]. These crystals may protrude from the freezing droplets into the continuous phase and, upon collision with another droplet which might be still in a liquid state due to the super-cooling phenomenon [91], they may puncture the emulsion film separating the neighboring drops, **Figure 5a**. The created drop-drop contacts induce partial or complete freezing of the second drop, because the protruding crystals act as nuclei for crystal growth within the volume of the second drop. The rapid freezing of the contacting drops precludes their complete merger which justifies the used term "partial coalescence". When many drop-drop contacts are formed upon storage or during shearing of the cooled emulsion, a 3D network of interconnected solid or semi-solid particles is formed and the disperse system increases very significantly its apparent viscosity, up to possible complete gelation [19,91,94].

The partial coalescence was studied in relation to several applications. In food systems, the partial coalescence might be a desired phenomenon, if a stable semi-solid dispersion with a desired plastic texture is to be produced (ice cream, whipped toppings, butter, etc.) [91,94]. In contrast, the partial coalescence should be avoided during the long-term storage (shelf-life) of sauces, creams, and other dairy products. The partial coalescence could appear also as a severe technological problem, if it occurs prematurely, as the rheological properties of the partially-coalesced dispersion may become inappropriate for its downstream processing. The particle coalescence is an important technological problem also when formulating stable alkane-in-water emulsions for energy storage and transportation (emulsions of phase-change-materials, PCMs), in which the latent heat of the liquid-solid phase transition of the alkane drops is used as energy reservoir [19].

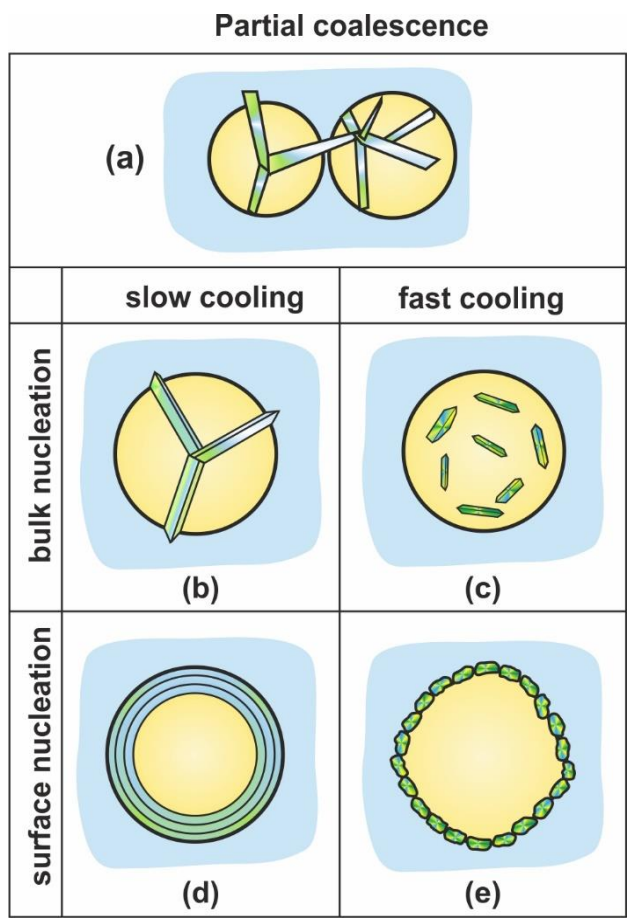

**Figure 5.** Schematic presentation of the effects of the crystal nucleation site (bulk or surface) and cooling rate (fast or slow) on the type of crystals formed and on the partial drop-drop coalescence in cooled emulsions. (a) Two droplets in close proximity – the growing lipid crystals may break the aqueous film between them, thus causing partial drop-drop coalescence. (b) Large crystals, protruding from the drop interior are formed in the case of bulk nucleation and at slow cooling rate. (c) Smaller crystals, which remain trapped inside the drop interior, are formed upon rapid cooling in bulk nucleation. (d) Large crystals grow from the drop surface inwards when slow cooling and surface nucleation are combined. (e) Crust of small crystallites is formed first in the case of surface nucleation and fast cooling. Due to the subsequent shrinkage of the oil upon freezing of the remaining oil in the drop interior, this crust can be distorted and crystal bulges can be formed on the surface of the freezing drop. Cases (b) and (e) are proposed in literature as promoting the partial coalescence, whereas cases (c) and (d) are considered as less efficient in inducing partial drop coalescence [95,96].

Therefore, the main factors for control of the partial coalescence have been studied widely, including the process parameters during emulsion preparation and storage, oil volume fraction and chemical composition, chemical composition of the continuous phase, drop size, and composition of the interfacial layer [92-94]. The results reported in the various papers are often in apparent contradictions with each other. These contradictions are briefly described below and a plausible unifying explanation, based on the role of surface freezing, is proposed.

Fat crystals protruding out of the surface of fatty droplets, which are able to penetrate into the aqueous phase and to break the emulsion films, were visualized in some experiments with large drops (15-20 μm), stabilized by polysorbate 20 (Tween 20) with 12 carbon atoms in the surfactant chain [95]. However, Moens et al. [96] did not confirm the occurrence of well-defined penetrating fat crystals in the partial coalescence phenomenon.

To explain the absence of such protruding crystals in their experiments, Moens et al. [96] proposed that the partial coalescence may involve the occurrence of local heterogeneities (corrugations) in the surface crust of frozen fat, instead of being caused by distinct fat crystals, protruding from the drop surface. These authors hypothesized that the surface crystalline crust may deviate from the initial spherical shape of the drop surface, due to the oil shrinkage upon drop freezing, resulting in formation of frozen fat bulges on the drop surface, **Figure 5e**, which may induce partial drop-drop coalescence in the systems studied by Moens et al. [96].

The effect of the cooling rate on the partial coalescence also remains controversial. At low cooling rate, the crystal growth process is favored over the nucleation process, resulting in smaller number of larger in size crystals. In contrast, at higher cooling rates, the nucleation process prevails and numerous smaller crystallites are formed [94,96]. Fredrick et al. [94] hypothesized that the larger crystals protrude over a longer distance into the aqueous phase, therefore enhancing the partial coalescence. Degner et al. [97] also concluded that slow freezing enhances the drop-drop aggregation. However, in their studies, Moens et al. [96,98] reported the opposite effect of the cooling rate – the emulsions were more stable at slower cooling rates, under otherwise equivalent conditions.

These apparent discrepancies could be resolved by considering the fact that, depending on the surfactant used for emulsion stabilization, the fat crystallization may nucleate at two different locations – in the drop interior or on the drop surface. Indeed, several studies revealed that the partial coalescence depends strongly on the melting temperature of the low-molecular-mass surfactants present [19,20,94,99]. The long-chain saturated monoglycerides form adsorption layers which may freeze before the freezing of the oily phase inside the drops, whereas the surfactants containing double bond (oleates) form fluid adsorption layers which cannot crystallize before the drop interior.

Thus we see that the effects of freezing of the surfactant adsorption layer and rate of emulsion cooling are interrelated. For fluid adsorption layers, the drop freezing nucleates in the drop interior and the fat crystals grow towards the drop periphery. Upon slow cooling, larger crystallites are formed which have greater protrusion into the aqueous phase, thus enhancing the partial coalescence, cf. **Figures 5a** and **5b** [99].

In contrast, for pre-solidified adsorption layers, the drop freezing starts from the drop surface and evolves towards the drop interior, templating lamella planes of the fat crystals with orientation parallel to the oil-water interface [20,95]. Here, the slow cooling should create fewer in number surface nuclei which slowly grow toward the drop interior, without breaking the neighboring emulsion films, **Figure 5d**. The fast cooling, however, should lead to dense coverage of the drop surface with a fragile crust of small crystallites, which may become corrugated with the subsequent freezing and shrinking of the remaining oil inside the droplets, thus inducing partial coalescence of the neighboring drops, **Figure 5e**.

In other words, the effect of the cooling rate could be opposite, depending on the type of surfactant used and on the ensuing location of the fat crystal nucleation, thus explaining the apparent discrepancies reported in literature.

The interfacial nucleation was related also to the observed increased stability toward partial coalescence in paraffin-in-water emulsions, studied as PCMs for thermal storage [19,21]. In these systems, the emulsifiers, which are in a liquid state around the temperatures of paraffin crystallization, were found to promote the partial coalescence [19,20].

We conclude that the surface freezing and the related surface nucleation of the fat/alkane crystals change dramatically the behavior of the emulsions in the process of partial coalescence. Further studies are needed to build up a univocal mechanistic explanation of the wide variety of behaviors observed and to reveal the convenient factors for process control.

*(c) Surface freezing in drop self-shaping and drop bursting phenomena.*

The crystallization of surfactant adsorption layers on the surface of oily drops turned out to be a crucial step in two other, recently discovered phenomena, observed in cooled emulsions – drop self-shaping [100*-102] and drop self-emulsification [103*,104], **Figure 6**. The main observations and the role of the surface freezing in these phenomena are briefly discussed below.

Emulsion drops from various long-chain organic substances, such as alkanes, alkenes, alcohols and their mixtures, were observed to exhibit a spectacular series of shape transformations upon emulsion cooling, but only when the oily drops are stabilized by surfactants with long saturated hydrophobic chains, e.g. with 16 or 18 carbon atoms [100*,101]. Two different mechanisms were proposed in the literature and are currently debated to explain this phenomenon, both of them starting with freezing of the surfactant adsorption layer on the oil drop surface, as a first step triggering a series of subsequent steps.

Denkov et al. [100*] assumed that the frozen surfactant layer serves as a template for ordering of the neighboring oil molecules, next to the drop surface, into molecular multilayers of the so-called "rotator" or "plastic" phases [105**,106]. These rotator phases have certain visco-elasto-plasticity which may overcome the capillary pressure of the droplets, which acts to preserve the spherical drop shape. Differential scanning calorimetry (DSC) was used to determine the thickness of the formed rotator phases which was found to vary between 2 and ca. 80 molecular layers, depending on the specific emulsion composition and drop size [106]. On the basis of this mechanism, a detailed theoretical model was proposed which explained the experimentally observed sequence of drop transformations [107,108].

In an alternative explanation, Guttman et al. [109*,110] proposed that the frozen surfactant monolayer could (*per se*) drive the drop shape transformations, due to ultra-low and even "transiently negative" interfacial tension which may occur in some non-equilibrium systems. Indeed, the Krafft point of $C_{18}$TAB surfactant, used in the experiments of Guttmann et al. [109*,110] is ≈ 36°C [111] which suggests that the experiments with hexadecane drops (melting temperature ≈ 18 °C) are performed at temperatures below the surfactant Krafft point, i.e. under non-equilibrium conditions.

It was shown later by these two groups that the shaped drops could be polymerized under appropriate conditions and that even sub-micrometer droplets can undergo self-shaping upon emulsion cooling [112,113].

The current brief review is not a place to debate about the different possible mechanisms of drop self-shaping. We note only that the interfacial tension was measured to be in the range between ca. 2 mN/m and 10 mN/m, viz. it was not ultra-low or negative for the surfactant-oil systems, studied by Denkov et al. [102]. Therefore, the mechanism, proposed by Guttmann et al. [109*,110], cannnot explain the results reported in Refs. [100*,101], though it might be operative for the experimental systems described in Ref. [109*,110]. Note that the two research groups always worked with different surfactant-oil systems and one cannot exclude the possibility that different mechanisms were operative in their studies.

In related studies, a drop "self-emulsification" phenomenon was observed upon cooling and heating of emulsions, stabilized by long-chain surfactants [103*,104]. Several detailed mechanisms were observed by optical microscopy to induce spontaneous drop fragmentation. All these mechanisms were related to the formation of dense adsorption layers which freeze upon emulsion cooling before the complete solidification of the oily drops. Some of these mechanisms are extremely efficient – after freezing and upon melting, the initial drops may burst into hundreds and thousands of sub-micrometer droplets without any mechanical input to the emulsion [103*,104].

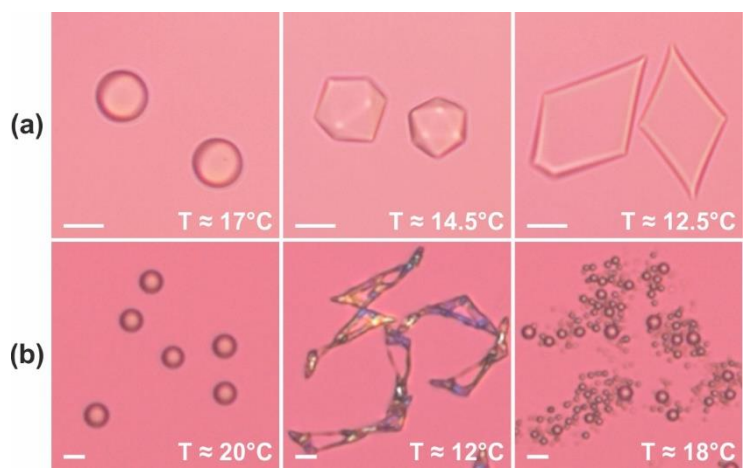

**Figure 6.** Microscopy images illustrating (a) drop self-shaping and (b) drop-bursting phenomena in cooled emulsions. (a) Hexadecane droplets, dispersed in 1.5 wt. % aqueous solution of Tween 40, have initially the characteristic spherical shape (left). Upon cooling, the drops transform into regular polyhedrons (in the middle) which gradually flatten until tetragonal platelets are formed (right). (b) From the initial six spherical hexadecane droplets (left), six frozen platelets are formed upon cooling (in the middle). Upon subsequent heating, the melting platelets disperse spontaneously into hundreds of smaller droplets (right). The hexadecane emulsion in (b) is stabilized by 1.5 wt.% Brij S20 nonionic surfactant. In all images the scale bars are 10 μm.

Related to these phenomena could be the study by Kovacik et al. [114] who performed temperature-dependent second harmonic sum frequency and light scattering measurements on alkane-in-water nanoemulsions, stabilized by $C_n$TAB. These authors concluded that, when the surfactant tail is longer than the alkane chain length, a transformation in the interfacial water ordering and freezing of the drop surface layer occur first, followed by solidification of the drop interior. It was not clarified whether the nano-droplets in [114] changed their shape when long-chain $C_n$TAB surfactant was used for emulsion stabilization – a possibility which is worthy to be explored in further studies.

### *(d) Saponin stabilized emulsions.*

Saponin stabilized emulsions were recently studied by several groups. In [115], a very good thermal stability of medium chain triglyceride (MCT) Miglyol-in-water emulsions, stabilized by *Yucca* saponin extract, was reported in the pH range between 5 and 9. In a separate study, the emulsions stabilized by *Quillaja* saponins were found to be sensitive to the variations in the pH and electrolyte concentration [116,117]. In both cases the emulsion stability was explained with electrostatic repulsion between the oily drops, without invoking any effects related to the surface viscoelasticity of the saponin adsorption layers. On the other hand, the studies of such layers at oil-water interfaces showed very high surface visco-elasticity for some saponin-oil combinations [64]. Peculiar in shape, non-spherical emulsion drops were observed in the respective emulsions [64]. The latter observations imply that the properties of some saponin-stabilized emulsion could be governed by the surface visco-elasticity of the respective adsorption layers – a possibility which should be checked in the future.

In a very interesting study [118*], Chen et al developed a procedure for formation of concentrated nanoemulsion of droplets with diameter of ≈ 100 nm, stabilized by *Quillaja* saponins. These nanoemulsions were aerated to obtain stable and highly viscoelastic foams with hierarchical internal structure. The peculiar properties of this system were explained with the high surface viscoelasticity of the saponin adsorption layers and with the strong adhesion of the nanodroplets to the bubble surfaces which resulted in Pickering type of foam stabilization.

Similarly, aerated emulsions were obtained with glycyrrhizic acid (another type of saponin) [119]. Nanofibrils and fibrillary network (hydrogel) were formed by the glycyrrhizic acid in the aqueous phase which blocked the Plateau channels and stabilized the foams at low temperature. The saponin hydrogel was destroyed at higher temperature, thus deteriorating foam stability.

The above overview evidences that further studies are needed to deepen our understanding of the factors controlling the surface phase transition in emulsions and the related complex phenomena.

## 4. Outlook.

The phase transitions on the surface of bubbles have been used for years to control dynamic properties of foams. Still, very little is known about the detailed structure of the adsorption layers in the most widely used surfactant-cosurfactant and saponin-containing systems with high surface viscoelasticity. Therefore, one may expect that the established experimental methods for surface structural analysis will be used in the next years to understand better the molecular structure-function relations in the formation of viscoelastic adsorption layers. The computer methods of molecular modelling (e.g. molecular dynamics) will be also a very useful tool to analyze the role and the relative contribution of the various types of intermolecular forces in the viscoelastic response of surfactant adsorption layers. Deeper understanding would allow the researchers and practitioners to design in more rational way specific surfactant mixtures with desired surface properties and controlled response to variations of pH, temperature, electrolytes and other stimuli. Interesting new approaches for foam control emerged in the last years, such as the use of photo-sensitive surfactants and fluorinated surfactants which both exhibit specific properties, very different from those of the known conventional surfactants.

The approaches for control of the surface phase transitions in emulsions are even less understood. Only recently the accumulated knowledge about the surface freezing of surfactant adsorption layers at the oil-water interfaces has been invoked to control the stability of regular emulsions. This area will certainly expand in the coming years due to the importance of emulsion partial coalescence in several industrial areas and to the discovery of the new intriguing phenomena of drop self-shaping and drop bursting in cooled emulsions, all strongly dependent on the process of surface freezing. Other new and unexpected phenomena could emerge when more intensive and systematic studies are performed of the role of surface phase transitions in these systems.

As evidenced in various studies, the surface phase transitions could trigger and could be involved in a complex interplay with other aggregation phenomena, such as surface templating and precipitation, surfactant-polymer complexation, and surfactant modification of particle hydrophobicity with subsequent surface or bulk particle aggregation/gelation. Such combined mechanisms were found to create ultra-stable foams and emulsions, with various potential applications, including their use as precursors for the fabrication of porous materials with hierarchical structure. The role of surface phase transitions has been recognized in these complex phenomena but has not been systematically studied yet. Therefore, understanding the role of surface phase transitions and revealing the main factors for their control would enhance our approaches for rational control over these complex phenomena with many potential applications.

**Acknowledgments:** This work was supported by the Bulgarian Ministry of Education and Science, under the National Research Programme "Healthy Foods for a Strong Bio-Economy and Quality of Life", approved by DCM #577/17.08.2018. The study falls under the umbrella of COST action CA17120 "Chemobrionics", funded by program Horizon 2020 of the EU.

*Using molecular dynamic simulations the authors show that the strong attraction between the saponin molecules in adsorption layers is due to the combined effect of long-range dipole-dipole interactions and short-range hydrogen bonds.*

*Strong surface complexation in mixed adsorption layers of Quillaja saponin (QS) with proteins and lipids is reported.*

## Graphical abstract

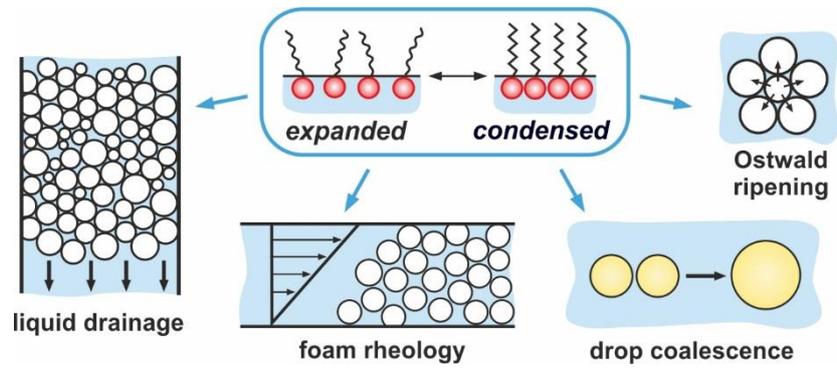